\documentstyle[12pt]{article}

\begin{document}

\baselineskip 16pt

\title{ WEAK-MEASUREMENT ELEMENTS OF REALITY}
\author{ Lev Vaidman}
\date{}
\maketitle

\begin{center}
{\small \em School of Physics and Astronomy \\
Raymond and Beverly Sackler Faculty of Exact Sciences \\
Tel Aviv University, Tel-Aviv 69978, Israel. \\}
\end{center}

\vspace{2cm}
\begin{abstract}
A brief review of the attempts to define ``elements of reality'' in
the framework of quantum theory is presented. It is noted that most
definitions of elements of reality have in common the feature to be a
definite outcome of some measurement. Elements of reality are extended
to pre- and post-selected systems and to measurements which fulfill
certain criteria of weakness of the coupling. Some features of the newly
introduced concepts are discussed. 
\end{abstract}


\newpage
\section{INTRODUCTION}
\label{int}

Experiments performed in laboratories of twenties century tell us that
the picture of reality we have from our everyday experience cannot be
a true representation of nature.  This leads us to develop a new
language which will be more adequate for the description of our
(quantum) world. The fact that our world is quantum seems indisputable
because the predictions of quantum theory have been confirmed with
incredible precision in all experiments which have been performed until
today.

There have been numerous attempts to describe quantum reality, but a
consensus has not been reached.  I do not have the pretension to give
here {\em the} correct definition of elements of reality in quantum
theory. I believe, however, that the concepts which I introduce are
helpful tools for developing intuition to see and to understand
quantum phenomena.

Einstein, Podolsky and Rosen (EPR) were the pioneers in their attempt
to define {\em elements of reality} in quantum theory. The difficulty
of considering elements of reality in the framework of quantum theory
which they discussed in their seminal paper\cite{EPR} led Bohr to
claim that the only elements of reality in quantum theory are results
of (quantum) measurements.
(This is, of course, an oversimplification. Max Jammer has illuminating
writings on this subject clarifying this important historical
issue\cite{Jamm1,Jamm2}.)
 The position of Bohr is
certainly consistent, but, I believe, is not very fruitful: it implies
that no reality
is associated with the quantum system between measurements.  This
approach refuses to consider various concepts which are helpful tools
to see the bizarre features of our quantum world.
The EPR definition of element of reality is:
\begin{quotation}
``If, without in any way disturbing the system, we can predict with
certainty (i.e. with probability equal to unity) the value of a physical
quantity, then there exists an element of physical reality corresponding to
this physical quantity.''
\end{quotation}
The key to understanding of the meaning of their definition is the
interpretation of the phrase ``without in any way disturbing the
system''. They assumed a strong version of relativistic causality,
i.e. that no operation in a space-like separated region can disturb
the system. I would  say that Bell \cite{Bell64} showed that
the existence of EPR elements of reality is inconsistent with
predictions of quantum theory. However there are many important
historical aspects of that issue as Max Jammer \cite{Jamm2} has
taught us.

The next definition of element of reality I want to bring here was
inspired by EPR but it makes no assumptions about relativistic
causality. This is the definition of element of reality due to Redhead
\cite{R}
  \begin{quotation}
``If we can predict with
certainty, or at any rate with probability one, the result of  measuring a
physical quantity at time $t$, then, at time $t$, there exists an element of
reality corresponding to this physical quantity and having value equal to
the predicted measurement result.''
  \end{quotation}
Although the definitions look very similar, they are very different
conceptually. The EPR element of reality is defined by the mere
possibility of finding the outcome ``without disturbing the system in
any way'', i.e., from some measurements, while according to Redhead,
there is an element of reality when we {\em know} the outcome of the
measurement without performing any other measurement.  For example, if we
consider an EPR pair of spin 1/2 particles (i.e. the two particles in
the singlet state) then, according to EPR, the spin components of each
particle in all directions are elements of reality since they can be
found from the spin measurement performed on the other particle (and the
assumption of no action at a distance ensures that this measurement
does not disturb the first particle). According to Redhead, however,
there is not any (local) element of reality for a spin variable of any
particle; we cannot predict the result of such measurement.

The three alternative definitions of element of reality which I
presented above are all related to definite results of measurements.
In the first (spirit of Bohr) these are the results of measurements
which were actually performed, in the last (Redhead) these are the
results of measurements which (if performed) are known with certainty,
and in the second (EPR) these are the results of measurements which we
can ascertain with certainty using other measurements (performed in a
space-like separated region). Thus, it is natural to adopt the
following principle:{\it  For any definite
  result of a  measurement there is corresponding element of reality.}

\section{ELEMENTS OF REALITY AS A DEFINITE SHIFT OF THE PROBABILITY DISTRIBUTION OF THE POINTER VARIABLE}
\label{ershift}

In order to understand the meaning of ``definite result of measurement''
we have to analyze the concept of quantum measurement. The standard
definition of quantum measurement is due to von Neumann \cite{vN}:
The measurement of a physical variable $A$ is described by the
Hamiltonian: 
\begin{equation}
  \label{ham}
 H = g(t) P A~~~~, 
\end{equation}
where $P$ is a canonical momentum conjugate to the pointer variable  $Q$ of
the measuring  device.
 The
function  $g(t)$ is nonzero only for a very short time
interval corresponding to the measurement, and is normalized so that
$\int g(t)dt=1$.
During  the time of  this 
impulsive measurement, the Hamiltonian (2) dominates the evolution of  the measured
system and the measuring device. Since $[A , H] =0$, the variable $A$ does
not change during the measuring interaction. The initial state of the
pointer variable is usually modeled by a  Gaussian centered at
zero:
\begin{equation}
  \label{phi-in}
\Phi _{in} (Q) =(\Delta ^2 \pi )^{-1/4} e^{ -{{Q ^2} /{2\Delta ^2}}}.
\end{equation}
The pointer is in the ``zero'' position before the measurement, i.e. its initial
probability distribution is 
\begin{equation}
  \label{prob}
prob(Q) = (\Delta ^2 \pi )^{-1/2} e^{ -{{Q ^2} /{\Delta ^2}}}.  
\end{equation}
 If the initial state of the system is an eigenstate 
$
|\Psi_1 \rangle = |a_i \rangle
$,
then after the interaction (1), the state of the system and the measuring device is:
\begin{equation}
  \label{mdstate}
(\Delta ^2 \pi )^{-1/4} |a_i \rangle e^{ -{{(Q-a_i)^2} /{2\Delta ^2}}}.  
\end{equation}
The probability distribution of the pointer variable, $(\Delta ^2 \pi )^{-1/2} e^{ -{{(Q-a_i)^2} /{\Delta ^2}}}$ remained unchanged in
its shape, but it is shifted by the eigenvalue $a_i$.
 This eigenvalue is
considered to be the element of reality. Thus we can translate the meaning
of ``definite result of quantum
measurement'' as {\em definite shift of the probability distribution of the
pointer variable}.
 I suggest  to take this property to be the  definition:
 {\em If we are certain that a  procedure for measuring  a certain variable
will lead to a definite shift of  the unchanged probability distribution
of the pointer, then there is an element of reality: the variable  equal to this shift.}

In an ideal measurement, the initial probability distribution of the
pointer is well localized around zero, and thus the final distribution
is well localized around the eigenvalue. Thus, the reading of the
pointer variable in the end of the measurement almost always yields  the
value of the shift (the eigenvalue of the variable). The
generalization I suggest in the above definition  is applicable
also to the situations in which  the initial probability
distribution of the pointer variable has a significant spread.
Although, in this case, the reading of the measuring device at the end
of the measurement is not definite, the shift of the distribution is.
In such a case, a measurement performed on a single system does not yield
the value of the shift (the element of reality), but such measurements performed on
large enough ensemble of identical systems yield the shift with any
desirable precision.

If the initial state of the system is a superposition 
$
|\Psi_1 \rangle = \Sigma \alpha_i |a_i \rangle
$,
then after the interaction (2) the state of the system and the measuring device is:
\begin{equation}
  \label{state}
(\Delta ^2 \pi )^{-1/4} \Sigma \alpha_i |a_i \rangle e^{ -{{(Q-a_i)^2} /{2\Delta ^2}}}.  
\end{equation}
The probability distribution of the pointer variable corresponding to the
state (5) is 
\begin{equation}
  \label{prob2}
prob(Q) =(\Delta ^2 \pi )^{-1/2} \Sigma |\alpha_i|^2  e^{ -{{(Q-a_i) ^2} /{\Delta ^2}}}.  
\end{equation}
In case of ideal measurement this is a weighted sum of the initial
probability distribution localized around various eigenvalues. It is
not a single shifted original probability distribution, and therefore,
according to our definition in this case there is no element of
reality for the value of the measured variable. (If one adds a
``collapse'' to one of the positions of the pointer, he ends up with
one of the eigenvalues. So, {\em a la} Bohr, after the measurement is
completed, there is an element of reality, but here we are looking for
an element of reality which can be predicted with certainty before the
measurement.)  In general, when the initial probability distribution
is not very well localized, the final distribution (6) is
very different from the original distribution. However, as I will show
in the next section, even in this case there is a well defined limit in
which the final distribution converges to the unchanged initial
distribution shifted by a well defined value. I will suggest to call
this well defined shift an element of reality of a new type.

\section{WEAK-MEASUREMENT ELEMENT OF REALITY}
\label{wmer}

I propose to consider the standard measuring procedure (1) in which we
weaken the interaction in such a way that the state of the quantum system
is not changed significantly during the interaction. Usually, the
measuring interaction leads to a very large uncertain change of the system
due to a large uncertainty of the variable $P$. Indeed, in the standard measurement
we require that the pointer shows zero before the measurement, i.e.,
$\Delta$ is very small for
the initial state of the measuring device (2). This requires  large uncertainty in $P$, and therefore the
Hamiltonian (1) causes a large uncertain change. I propose to take the initial
state of the measuring device in which  $\langle P \rangle = 0$ and
the uncertainty in $P$ is small, and I will
show that this is enough to ensure that the pointer probability distribution
after the measuring interaction is essentially equal to the shifted initial
distribution. In this case  the interaction Hamiltonian (1)  is small  and
this is why we call such procedure  a {\em weak
  measurement}\cite{AV-weak}.

The limit of weak measurement corresponds to  $\Delta \gg a_i$ for all  eigenvalues
$a_i$. Then,  we can  perform the Taylor expansion of the sum (6) around
$Q=0$ up to
the first order and rewrite the probability distribution of the
pointer in the following way:
\begin{eqnarray}
  \label{prob2}
 prob(Q) =(\Delta ^2 \pi )^{-1/2} \Sigma |\alpha_i|^2 e^{ -{{(Q-a_i)^2}
/{\Delta ^2}}} =~~~~~~~~~~~~~~~~~~~~~~~~~~~~~~~~~\nonumber\\
(\Delta ^2 \pi )^{-1/2} \Sigma |\alpha_i|^2 (1
-{{(Q-a_i) ^2} /{\Delta ^2}}) = (\Delta ^2 \pi )^{-1/2} e^{
 -{{(Q-\Sigma|\alpha_i|^2a_i) ^2} /{\Delta ^2}}} . 
\end{eqnarray}
But this is exactly the initial distribution shifted by the value
$\Sigma|\alpha_i|^2a_i$. Thus we will say that there is here a
{\em weak-measurement element of reality} $A_w = \Sigma|\alpha_i|^2a_i$.

The mathematical expression of this weak-measurement of reality is not
something new. This is the expectation value: $A_w =
\Sigma|\alpha_i|^2a_i =\langle \Psi |A|\Psi\rangle$. The weak value is
obtained from statistical analysis of the readings of the measuring
devices of the measurements on an ensemble of identical quantum
systems. But it is different conceptually from the standard definition
of expectation value which is a mathematical concept defined from the
statistical analysis of the {\em ideal} measurements of the variable
$A$ all of which yield one of the eigenvalues $a_i$.

I have showed that expectation values fall under the definition of the
elements of reality as a definite shift of unchanged  probability
distribution of the pointer variable in the limit of weak-coupling
measurement ($\Delta$ large). The advantage of this definition is that it
is applicable for any quantum system in any pure or mixed state and
for any quantum variable of this system. The disadvantage is that
usually we need an ensemble of identical systems in order to find the
expectation values of their quantum variables with  good
precision. It is important to mention that sometimes the {\em weak}
measurement need not be too weak and the expectation value can be
found with  relatively good precision from a single such
measurement. This is the case when the uncertainty of $A$ is
small. The latter, in particular, is a generic property of ``average''
variables of large composite systems.

\section{ ELEMENTS OF REALITY OF PRE- AND POST-SELECTED SYSTEMS}
\label{erpps}

The concept of the weak-measurement element of reality yields novel
results when considered on pre- and post-selected systems. But  the
ideal-measurement element of reality of the pre- and post-selected system
also have novel features and I will consider them first.

For the pre- and post-selected systems it is fruitful to consider a
modification of Redhead's definition when I replace ``predict'' by
``infer''\cite{V-prl,V-er}.
\begin{quotation}
``If we can infer with
certainty, or at any rate with probability one, the result of  measuring a
physical quantity at time $t$, then at time $t$, there exists an element of
reality corresponding to this physical quantity and having a value equal to
the predicted measurement result.''  
\end{quotation}
Essentially, Redhead's definition says that $A=a$ if and only if the
system is in the appropriate eigenstate   or mixture of
such eigenstates. In pre-and post-selected situations we might have $A=a$
even if the system  is not in an eigenstate of $A$; we obtain the inference
both from the preparation and from the post-selection measurement.

Elements
of reality in the pre- and post-selected situations might be very peculiar.
One such example is  a single  particle inside three boxes $A, B$ and $C$, with  two
elements of reality: ``the particle is in box 1'' and  ``the particle is in
box 2''. This is the case with the pre-selection of the state of the particle $|\Psi_1 \rangle = 1/\sqrt 3 ~(|A\rangle + |B\rangle +
|C\rangle)$ which was found later in the state $|\Psi_2 \rangle = 1/\sqrt 3
   ~(|A\rangle + |B\rangle -|C\rangle)$. If, in the intermediate time it was
searched for in  box $A$ it has to be found there with probability one, and if, instead,  it was
searched for in  box $B$,  it has to be found there too with probability one. (Indeed, not finding
the particle in box $A$ would project the initial state $|\Psi_1 \rangle$ onto $ 1/\sqrt 2 ~( |B\rangle +
|C\rangle)$ which is orthogonal to the final state $|\Psi_2\rangle$.)

 This
example shows that the ``product rule'' does not hold for elements of reality
of pre- and post-selected systems. Indeed $\Pi_A =1$ is an element of
reality and  $\Pi_B =1$ is an element of reality, but  $\Pi_A \Pi_B =1$  is not an element of
reality. In fact, $\Pi_A \Pi_B =0$ is an element of reality. The
meaning of this equation is a  trivial point that the probability to
find the particle both in $A$ and $B$ is equal  to zero.
The
``sum rule'' does not hold either. Indeed, there is no element of
reality $\Pi_A + \Pi_B =2$. In fact,  there is no any element of
reality for the value of the sum $\Pi_A +
\Pi_B$. This means that when we perform a
measurement which tells us that the particle is inside one of the
boxes $A$ or $B$, but without telling in which one, the probability to
find it is neither zero nor one. (The probability is equal 2/3, but
this does not correspond to any element of reality.) Note, however, that
there is an element of reality for the sum of the three projection
operators: $\Pi_A + \Pi_B + \Pi_C=1$; clearly, the measurement testing
the existence of  the particle in the three boxes will say yes with
probability one.
  
The elements of reality for pre- and post-selected quantum systems
have unusual and counterintuitive properties. But, maybe this is not
because of the illness of their definition, but due to bizarre features
of quantum systems which goes against the intuition built during
thousands of years, when the results of quantum experiments were not known.

\section{WEAK-MEASUREMENT ELEMENTS OF REALITY OF PRE- AND POST-SELECTED SYSTEMS}
\label{wmerpps}

The next natural step is to consider the limit of weak measurements
performed on pre- and post-selected quantum systems. Again we consider
the measuring Hamiltonian (1), the initial state of the measuring
device (2) and the limit of large $\Delta$. In general, this will lead
to a very low precision of the measurement so we consider an ensemble
of identical systems with such measurements. The difference here from
the weak measurement of Sec. 3 is that now, before reading the
outcomes of the measuring devices, we post-select a certain state of
the system and discard the readings of measuring devices corresponding
to the systems for which the post-selection was not successful. 

I will not repeat here the calculations, they can be found in
Ref.(7). The above procedure is called {\em weak measurement} and it is
indeed converges to a well defined value. At the limit of large
$\Delta$, the probability distribution of the final state of the measuring
device converges to the initial distribution shifted by the real part
of the {\em weak value}  of the variable $A$:
\begin{equation}
  \label{a-weak}
A_w = {{\langle \Psi_2|A|\Psi_1\rangle} \over {\langle
    \Psi_2|\Psi_1\rangle}}.  
\end{equation}
Our definition of elements of reality, i.e., a definite shift of the
probability distribution of the pointer variable yields for pre-and
post-selected systems the weak value (8). Even the imaginary part of
the weak value falls under this definition, but only for a particular case
of the Gaussian as the initial measuring device state. In order to see
that we have to consider, instead of the pointer position
distribution, the distribution of the conjugate momentum. It turns
out\cite{AV-weak} that the original Gaussian in $P$ does not change
its shape (again, in the limit of  a large $\Delta$) and is shifted by
the value $Im(A_w)$.

The weak-measurement elements of reality of pre-selected only systems,
the expectation values, are a particular case of the weak
values. Indeed, we can consider a future measurement which tests that
we are in the initial state $|\Psi_1\rangle$. The weak measurement, by
assumption, does not change the state of the system significantly, and
therefore, this future measurement {\em must} yield the state
$|\Psi_1\rangle$. But then, the definition of weak value (8) yields
the expectation value.

The advantage of weak-measurement elements of reality is that they are well defined for any situation and for all variables. It also have
some classical type features. The ``sum rule'' holds. Indeed, if $C =
A + B$ then  $C_w =
A_w + B_w$. Therefore, if $A_w =a$ is a weak-measurement element of
reality and  $B_w = b$ is a weak-measurement element of
reality, then $(A+B)_w=a +b$ is also a weak-measurement element of
reality. The ``product rule'', however, does not hold. From  $C =
A  B$ does not follow that  $C_w =
A_w  B_w$.     

The main disadvantage of weak-measurement elements of reality is again that usually we need an ensemble of
identical pre- and post-selected systems in order to find the weak
values of their quantum variables with a good precision. However, there
are certain important cases in which the {\em weak} measurement need
not be too weak and the ``weak'' value can be found with a relatively
good precision from a single such measurement. For example if a spin
$N$ particle is prepared in the state $S_x=N$ and later found in the
state $S_y=N$, then, at the intermediate time the weak value of the spin component in the direction $\hat{\xi}$ which bisects
$\hat{x}$ and $\hat{y}$ is  larger than $N$. Indeed, $(S_\xi)_w
=\sqrt{2} N$. An experimenter  can repeatedly see this
``forbidden'' value in a standard measurement with precision of order
$\sqrt{N}$. Note, however, that for any ``unusual''
weak values the probability to obtain the required result of the
post-selection is extremely small. (In the last example this
probability is equal $2^{-N}$.)

The concept of weak-measurement elements of reality is a generalization  of the
usual concept of the (strong-measurement) element of reality. Indeed, if
we know with certainty that a strong measurement of $A$ will yield
$A=a$ with probability one, then we know that the weak measurement
will also
yield $A_w= a$ \cite{foot2}.
 Thus, all (strong) elements of
reality are also weak elements of reality. The class of weak elements
of reality is much wider; it is defined for all variables for any
realizable pre- and post-selected situation (as well as for
the pre-selection only situation). In contrast, the strong elements of
reality are defined only for some variables in each situation and
sometimes they do not exist at all. (There is no any {\em local} element of
reality for  spin variables of an EPR pair.)

Let us analyze the example of the three boxes we have introduced
above. Since we know several (ideal-measurement) elements of reality
we can immediately write down corresponding   weak-measurement elements of reality for the
discussed pre- and post-selected particle:
\begin{equation}
  \label{psi1}
(\Pi_A)_w =1,~~ (\Pi_B)_w =1,~~ (\Pi_A+ \Pi_B +\Pi_C)_w =1 . 
\end{equation}
Now using the sum rule we obtain another  weak-measurement element of
reality:
\begin{equation}
  \label{psi2}
(\Pi_C)_w =   (\Pi_A+ \Pi_B +\Pi_C)_w- (\Pi_A)_w -(\Pi_B)_w = -1 . 
\end{equation}
To say that there is a ``reality'' of having $-1$ particle in a box
sounds paradoxical. However, when we test this reality weakly this is
what we see. We cannot see this ``reality'' for one particle because
the uncertainty  of the appropriate weak measurement has to be much
larger than 1, but if we have a large number of such pre- and
post-selected particles in the three boxes, then a realistic
measurement of the pressure in the boxes will yield $p$ for boxes $A$
and $B$ and the negative pressure, $-p$, for the measurement in the
box $C$.
Note, however, that the probability to obtain in the post-selection
measurement the state $|\Psi_2\rangle$ for a macroscopic number of particle is 
extremely small.

\section{CONCLUSION}
\label{CONC}

The name ``element of reality'' suggests an ontological
meaning. However, historically, and in the present paper, the element
of reality is an epistemological concept. It is better to have 
ontological elements of reality, but there are  severe  difficulties
in constructing them. Probably, the most serious attempt in this
direction is the causal interpretation\cite{Bohm} which introduces
``real'' point-like particles moving according to a simple (but nonlocal)
law. However, it seems that we are forced to accept that the total quantum
wave in the causal interpretation has also ontological status\cite{AV-Bohm}. 
I feel that if we do consider the quantum wave function of the universe as
an ontological reality, we need not  add anything else, e.g. Bohmian
particles. I am perfectly ready to accept that the reality of our
physical universe {\em is} its wave function. But since that reality
is very far from what we experience, I think it is fruitful to define
epistemological reality as it has been done here.

I certainly see a deficiency of weak-measurement elements of reality defined above
in the situations in which they cannot be measured on a single system.
Still, I do not think that the fact that a weak value cannot be
measured on a single system prevents it from being a ``reality''. We
know that the measuring device shifts its pointer exactly according to
the weak value, even though we cannot find it because of the large
uncertainty of the pointer position. We can verify this knowledge
performing measurements on an ensemble.  In the cases when weak values
can be found with good precision on a single system, the concept of
weak-measurement elements of reality is fully justified. In fact, the
word ``weak'' is not exactly appropriate, since the measurements in
question are the usual one. (However, they are not ``ideal'' in the von Neumann
sense, since the measuring Hamiltonian has to be bounded.)

 Another
example of measurements which are ``good measurements'' in the sense
that they yield a measured quantity with a good precision and which
yield weak-measurement elements of reality are {\em protective
measurements}. For pre-selected protected systems these elements of
reality are again
expectation values\cite{AV-prot}. It has been shown that even
the two-state vector can be protected and that the weak values can be
measured on any single (appropriately protected) system with a good
precision\cite{AV-nyas}. Finally,  it has been
shown recently\cite{AMPTV}, that adiabatic measurements performed on decaying systems which
were post-selected not to decay, yield one of the (non-trivial, i.e.
not expectation value) weak values even without specific pre- and
post-selection, manifesting again the physical meaning of the weak values.

The concept of elements of reality for pre- and post-selected systems
fits very well the many-worlds interpretation\cite{mwi-eve} which I
believe is the best available interpretation today\cite{mwi-V}.  It
answers in a very convincing way the following difficulty. Consider a
present moment of time $t$. We can assign a list of elements of
reality and weak-measurement elements of reality based on the quantum
state at that time. It includes the eigenvalues of variables for which
the quantum state is an eigenstate and expectation values for all
variables. In particular, let us assume that a spin $N$ particle has
an element of reality $S_x =N$, and consequently it has no element of
reality regarding the value of $S_y$, but it has weak-measurement
elements of reality $(S_y)_w =0$. Let us assume that at a later time
the spin in the $y$ direction was measured and the (very improbable
for large $N$) result $S_y =N$ was obtained. If now we assign elements
of reality for the time $t$, we see that the list is different. Indeed, we
have to add some elements of reality, e.g., $S_y =N$ in our case, and we
have to change some weak-measurement elements of reality, e.g.,
$(S_y)_w =N$. How can we associate different elements of reality for
the same moment of time? The answer is natural in the framework of the
many worlds interpretation. We discuss here epistemological element of
reality of  conscious beings. Before the measurement of $S_y$ we
considered one world corresponding to a certain quantum state and in
this world we had certain elements of reality. The measurement of
$S_y$ generated $2N +1$ new worlds corresponding to different outcomes
of the measurement. The conscious beings (the experimenters) in these
different worlds have, not surprisingly, different sets of elements of reality.

  The research was supported in part by grant 614/95 of
 the Basic Research Foundation (administered by the Israel Academy
of Sciences and Humanities).


\begin{thebibliography}{99} 


\bibitem{EPR}
A. Einstein, B. Podolsky, and N. Rosen, {\it Phys.  Rev.} {\bf 47}, 777 (1935)
\bibitem{Jamm1}
M. Jammer, {\it The Conceptual Development of Quantum  Mechanics,}
(McGraw-Hill, New York, 1966).
\bibitem{Jamm2}
M. Jammer, {\it The Philosophy of Quantum Mechanics,} (Wiley, New
  York, 1974).
\bibitem{Bell64}
J.S. Bell, {\it Physics} {\bf 1}, 195 (1964).
\bibitem{R}
 {M. Redhead, {\it Incompleteness, Nonlocality, and Realism} (Clarendon,
Oxford, 1987) p. 72.}
\bibitem{vN}
 J. von Neumann,  {\it Mathematical Foundations of Quantum Theory},
  (Princeton, University Press, New Jersey (1983).
\bibitem{AV-weak}
 Y.Aharonov and L. Vaidman, {\it Phys.  Rev.}   {\bf A 41}, 11 (1990).
\bibitem{V-prl}
 L. Vaidman
{\it Phys.  Rev.  Lett.} {\bf 70}, 3369-72 (1993).
\bibitem{V-er}
{L. Vaidman, in {\it Symposium on the Foundations  of Modern Physics},
 {P.J. Lahti, P.~Bush, and P. Mittelstaedt (eds.), World
Scientific, 406-417 (1993).}}
\bibitem{foot2}
 {It is interesting that for {\em dichotomic variables} the
following is true: If the weak value is equal to one of the
eigenvalues, i.e. $A_w = a_i$ is the weak-measurement elements of
reality, then $A=a_i$ is an element of reality in the strong sense,
ideal measurement of $A$ yields $a_i$ with probability one \cite{AV-gwm}.}
\bibitem{AV-gwm}
Y. Aharonov and L. Vaidman
{\it Jour. Phys.} {\bf A 24}, 2315-28 (1991).
 \bibitem{Bohm}
 {D. Bohm, {\it  Phys.  Rev.} {\bf  85}, 97 (1952).}
\bibitem{AV-Bohm}
{ Y.  Aharonov and L. Vaidman, in 
'Bohmian mechanics and quantum theory: an appraisal,' J.T. Cushing, A.
Fine, and S. Goldstein. eds., (Kluwer, 1996)e-board: quant-ph/9511005}

\bibitem{AV-prot}Y. Aharonov and L. Vaidman, {\it Phys. Lett.} {\bf A178}, 38
(1993).

\bibitem{AV-nyas}  Y.Aharonov and L. Vaidman, {\it Ann. NY Acad. Sci.}
  {\bf 755}, 361 (1995).

\bibitem{AMPTV}  Y.Aharonov, S. Massar, S. Popescu, J. Tollaksen, and
  L. Vaidman, TAUP 2315-96.
\bibitem{mwi-eve}
{ Everett, H. (1957), 
{\it  Rev. Mod. Phys.}  {\bf 29}, pp. 454-462.}
\bibitem{mwi-V}
 {Vaidman, L. (1993), ``On  Schizophrenic Experiences of the Neutron
   or why We should Believe in the Many-Worlds Interpretation of
   Quantum Theory'', Tel-Aviv University preprint, TAUP 2058-93.}

\end{thebibliography}
\end{document}